\documentclass[prb,aps, twocolumn, amsmath,amssymb,superscriptaddress,longbibliography]{revtex4-1}

\usepackage{graphicx}
\usepackage{dcolumn}
\usepackage{bm}
\usepackage[sort&compress]{natbib}
\usepackage{xspace}
\usepackage{amsmath}
\usepackage{epstopdf}
\usepackage{amssymb}
\usepackage{epsfig}
\usepackage{color}

\begin{document}

\title{Deterministic implementation of a bright, on-demand single photon source with near-unity indistinguishability via quantum dot imaging}

\affiliation{Technische Physik and Wilhelm Conrad R\"ontgen Research Center for Complex Material Systems, Physikalisches Institut,
Universit\"at W\"urzburg, Am Hubland, D-97074 W\"urzburg, Germany}
\affiliation{Center for Nanoscale Science and Technology, National Institute of Standards and Technology, Gaithersburg, MD 20899, USA}
\affiliation{Present address: Hefei National Laboratory for Physical Sciences at the Microscale and Department of Modern Physics,
$\&$ CAS Center for Excellence and Synergetic Innovation Center in Quantum Information and Quantum Physics,
University of Science and Technology of China, Hefei, Anhui 230026, China}
\affiliation{Maryland NanoCenter, University of Maryland, College Park, MD 20742, USA}
\affiliation{School of Physics, Sun-Yat Sen University, Guangzhou, 510275, China}
\affiliation{SUPA, School of Physics and Astronomy, University of St Andrews, St Andrews, KY16 9SS, United Kingdom}

\author{Yu-Ming He}
\affiliation{Technische Physik and Wilhelm Conrad R\"ontgen Research Center for Complex Material Systems, Physikalisches Institut,
Universit\"at W\"urzburg, Am Hubland, D-97074 W\"urzburg, Germany}
\affiliation{Present address: Hefei National Laboratory for Physical Sciences at the Microscale and Department of Modern Physics,
$\&$ CAS Center for Excellence and Synergetic Innovation Center in Quantum Information and Quantum Physics,
University of Science and Technology of China, Hefei, Anhui 230026, China}

\author{Jin Liu}\email{liujin23@mail.sysu.edu.cn}
\affiliation{Center for Nanoscale Science and Technology, National Institute of Standards and Technology, Gaithersburg, MD 20899, USA}
\affiliation{Maryland NanoCenter, University of Maryland, College Park, MD 20742, USA}
\affiliation{School of Physics, Sun-Yat Sen University, Guangzhou, 510275, China}

\author{Sebastian Maier}
\affiliation{Technische Physik and Wilhelm Conrad R\"ontgen Research Center for Complex Material Systems, Physikalisches Institut,
Universit\"at W\"urzburg, Am Hubland, D-97074 W\"urzburg, Germany}

\author{Monika Emmerling}
\affiliation{Technische Physik and Wilhelm Conrad R\"ontgen Research Center for Complex Material Systems, Physikalisches Institut,
Universit\"at W\"urzburg, Am Hubland, D-97074 W\"urzburg, Germany}

\author{Stefan Gerhardt}
\affiliation{Technische Physik and Wilhelm Conrad R\"ontgen Research Center for Complex Material Systems, Physikalisches Institut,
Universit\"at W\"urzburg, Am Hubland, D-97074 W\"urzburg, Germany}

\author{Marcelo Davan\c co}
\affiliation{Center for Nanoscale Science and Technology, National Institute of Standards and Technology, Gaithersburg, MD 20899, USA}

\author{Kartik Srinivasan}
\affiliation{Center for Nanoscale Science and Technology, National Institute of Standards and Technology, Gaithersburg, MD 20899, USA}

\author{Christian Schneider}\email{christian.schneider@physik.uni-wuerzburg.de} 
\affiliation{Technische Physik and Wilhelm Conrad R\"ontgen Research Center for Complex Material Systems, Physikalisches Institut,
Universit\"at W\"urzburg, Am Hubland, D-97074 W\"urzburg, Germany}

\author{Sven H\"ofling}
\affiliation{Technische Physik and Wilhelm Conrad R\"ontgen Research Center for Complex Material Systems, Physikalisches Institut,
Universit\"at W\"urzburg, Am Hubland, D-97074 W\"urzburg, Germany}
\affiliation{SUPA, School of Physics and Astronomy, University of St Andrews, St Andrews, KY16 9SS, United Kingdom}

\date \today


\begin{abstract}
Deterministic techniques enabling the implementation and engineering of bright and coherent solid-state quantum light sources are key for the reliable realization of a next generation of quantum devices. Such a technology, at best, should allow one to significantly scale up the number of implemented devices within a given processing time. In this work, we discuss a possible technology platform for such a scaling procedure, relying on the application of nanoscale quantum dot imaging to the pillar microcavity architecture, which promises to combine very high photon extraction efficiency and indistinguishability. We discuss the alignment technology in detail, and present the optical characterization of a selected device which features a strongly Purcell-enhanced emission output. This device, which yields an extraction efficiency of $\eta=(49\pm4)~\%$, facilitates the emission of photons with $(94\pm2.7)~\%$ indistinguishability.
\end{abstract}

\maketitle

\section{Introduction}

More than 15 years after the observation of photon antibunching from a single quantum dot~\cite{michler2009single}, the engineering of bright solid-state single photon sources is still a topic of major interest. Establishing optical quantum networks on- and off-chip ~\cite{yao09,Hoang12,Dietrich16}, schemes for quantum teleportation~\cite{Nilsson-NatPhot13, Gao-NatCom13} and most importantly, the implementation of quantum repeater networks~\cite{Jones2016}, crucially relies on such quantum devices. While alternative platforms are being investigated, epitaxially grown single quantum dots (QDs) hold great promise due to their large optical oscillator strength, enabling single photon operation frequencies into the GHz range.
A modern single photon source which can be of use in any of the above mentioned applications has to fulfill a variety of prerequisites. It has to be operated on-demand, for which a strictly resonant pulsed excitation process has proven to be very suitable~\cite{He2013}. Sufficiently large photon count rates need to be achieved, which has mostly been realized by embedding single QDs in photonic architectures, including antennas and microcavities~\cite{Moreau2001,Pelton2002}, on-chip waveguides~\cite{arcari14}, solid immersion lenses \cite{Maier2014, Gschrey2015}, gratings~\cite{Sapienza2015}, and vertical nanowires~\cite{Claudon2010a,Heinrich2010,Reimer2012}. Via such architectures, QD-based single photon sources with extraction efficiencies in excess of 70 $\%$ ~\cite{Claudon2010a, Unsleber2016, Gazzano2013} have been reported. Applications relying on quantum interference (such as a Bell state measurement in an entanglement swap) also demand the highest degree of indistinguishability of the emitted photons. Recent efforts have outlined clear strategies to improve it. Indistinguishable photons share all characteristics, including polarization and color, and their quantum interference is typically probed in a single photon interference experiment. Thus, the characteristic destructive quantum interference between photons leaving separate output ports on a beam-splitter can only be established if the single photons impinge on the splitter at exactly the same time, and if their wave-packet overlap equals unity. Equal timing requires an excitation technique which minimizes the time jitter of the emission event, while maximizing the wave-packet overlap requires close to Fourier transform-limited photons. It has been recognized that resonance fluorescence excitation is the most suitable configuration to simultaneously minimize time jittering and maximize coherence ~\cite{He2013}. In addition, spontaneous emission enhancement via a microcavity resonance can improve the indistinguishability of the emitted photons ~\cite{Unsleber2015c, Varoutsis2005, Ding2016, Somaschi2016}.

Likely, the most challenging aspect in the implementation of QD-based single photon devices relying on advanced photonic structures stems from the random nature of the QD nucleation process. A genuinely scalable approach to embed quantum emitters in photonic devices can only be based on ordered QD arrays \cite{Schneider2009} with full control over the spectral characteristics.  While great progress has been made into this direction, in most cases, the emission properties of such positioned QDs are still compromised by the fabrication technology.  As a result, techniques to deterministically embed a single, pre-selected quantum emitter in a photonic device have been developed \cite{Gschrey2015, Badolato2005, Dousse2008, Thon2009, Iwamoto2016, Kojima2013}. However, most of these techniques rely on the subsequent identification of individual QDs using scanning techniques (such as confocal microscopy or cathodoluminescence) that can have low throughput.
Here, we demonstrate a deterministic implementation of single QDs in micropillar cavities based on a nanoscale QD imaging technology~\cite{Sapienza2015}. This technology can, in principle, be more scalable since the identification of a number of QD positions can be carried out in a single shot with nanometer accuracy, and the multiplexed nature of wide-field illumination and camera detection enables rapid mapping of an entire sample. We then demonstrate, for a selected device, that our approach can yield single photon extraction efficiencies close to the state-of-the-art in the field, and more importantly, can be operated in pulsed resonance fluorescence, resulting in single photon streams with near-unity indistinguishability.

\section{Deterministic fabrication of micropillar cavities}

\begin{figure}[t]
\includegraphics[width=0.75\linewidth]{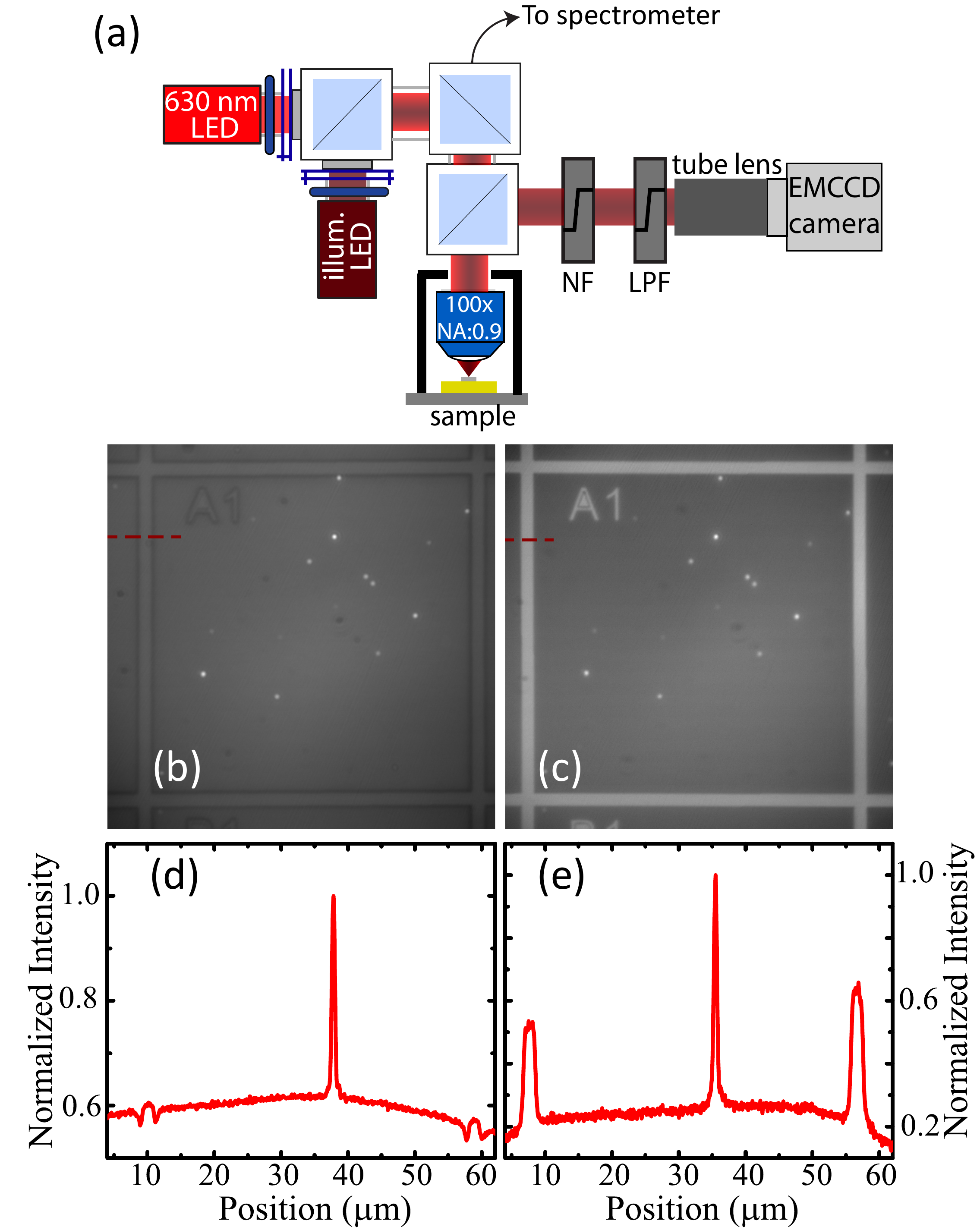}
\caption{(a) Schematic of the photoluminescence imaging setup used for determining QD locations en route to deterministic micropillar single photon source fabrication.  Illumination of the alignment marks is done by a 730~nm LED, while removal of unwanted light entering the EMCCD camera is done through two notch filters (NFs) that block wavelengths between 810~nm and 880~nm, and a long pass filter (LPF) that removes wavelengths below 700~nm. (b) Image acquired using the original photoluminescence imaging setup configuration described in Refs.~\cite{Sapienza2015,Liu2016}, in which the illumination LED is at 940~nm.  (c) Image acquired using the modified setup depicted in (a). (d) Horizontal line cut through the image from (b), along the dashed red line.  While the contrast between the QD emission and background level is high (central peak), the alignment mark contrast is limited. (e) Horizontal line cut through the image from (c), along the dashed red line.  The modified imaging setup yields good contrast between the QD emission and background level as well as between the alignment marks and background signal level.  The alignment mark separation is 52~$\mu$m in (b)-(e).}
\label{Fig1}
\end{figure}

We take advantage of the bichromatic fluorescence imaging method developed in Ref.~\cite{Sapienza2015} to locate the spatial positions of single QDs with respect to pre-defined metallic alignment marks.  In this approach, schematically depicted in Fig.~\ref{Fig1}(a), a 630~nm LED is used to excite all of the QDs within the system's field of view (typically $\approx~60~\mu$m~$\times$~60~$\mu$m), while a long wavelength LED (typically near the QD emission band in the 900~nm range) simultaneously illuminates the sample.  Emitted light from the QDs and reflected light off the sample are directed through one or more filters to reject light from the short wavelength LED before going into a sensitive electron-multiplied charge-coupled device (EMCCD) camera. Significant improvement in the performance of this approach has been reported in Ref.~\cite{Liu2016}, where use of a high numerical aperture objective within the sample's cryogenic environment resulted in a reduced image acquisition time (now 1~s) and lower uncertainties in the localization of the QDs and alignment mark centers (factors of 10$\times$ and 4.7$\times$, respectively).

However, a straightforward application of the setup demonstrated in Refs.~\cite{Sapienza2015,Liu2016} to samples with distributed Bragg reflectors (DBRs) yields poor results.  For example, Fig.~\ref{Fig1}(b) shows an image acquired by applying the setup from Ref.~\cite{Liu2016} to our sample, which consists of 25.5 (15) $\lambda/4$-thick AlAs/GaAs mirror pairs which form the lower (upper) DBR, with metallic alignment marks fabricated on the sample surface by electron-beam (e-beam) lithography, thermal evaporation, and a lift-off process. While the signal-to-noise ratio of the QDs in Fig.~\ref{Fig1}(b) is similar to what has been achieved in previous works, the alignment marks show very low contrast. The basic issue is that the contrast relies on a difference in reflectivity between the alignment marks and the sample at the illumination wavelength of 940~nm.  For samples without DBR stacks, this contrast is significant due to the large difference in reflectivity between the Au marks ($>95~\%$ reflectivity) and the GaAs surface ($\approx$~30~$\%$ reflectivity) at 940~nm.  However, the DBR reflectivity is specifically engineered to be high (larger than that of Au), so that little contrast is observed when illuminating the sample at this wavelength.  In fact, the ability to discern the alignment marks in Fig.~\ref{Fig1}(b) is primarily due to the $\approx$100~nm difference in height between the alignment marks and the sample surface, which results in the shadow-like dark regions surrounding the marks, as evident from a line cut through the image, shown in Fig.~\ref{Fig1}(d).

Since Au is a spectrally broadband reflector but the DBR mirror has a narrower spectral bandwidth, we can adjust the illumination wavelength to regain contrast in the image.  An important constraint is to continue to image the QD emission at the same time, while suppressing the QD excitation LED and unwanted emission from the sample (e.g., from the GaAs band edge and the wetting layer states).  The most straightforward solution would be to move the illumination LED to a longer wavelength, where the DBR is no longer highly reflective, but the DBR bandwidth is sufficiently spectrally broad so as to require wavelengths outside of the EMCCD detection range.  Instead, we settle on an illumination wavelength of 730~nm.  While in principle this 730~nm light also pumps the QDs, the intensity we use for alignment mark imaging is orders of magnitude lower than the intensity of the 630~nm light used to excite the QDs.  We then use a series of filters to remove unwanted light entering the EMCCD, namely a 700~nm long pass filter to remove the reflected 630~nm pump, and two 800~nm band notch filters to remove the GaAs band edge emission near 830 nm and the QD wetting layer emission near 850 nm. The resulting image is shown in Fig.~\ref{Fig1}(c), where the contrast in the alignment marks and signal-to-noise in the QD emission (Fig.~\ref{Fig1}(d)), and the corresponding total position uncertainty ($<10$~nm), are similar to those achieved in Ref.~\cite{Liu2016}.

\begin{figure}[t]
\includegraphics[width=0.95\linewidth]{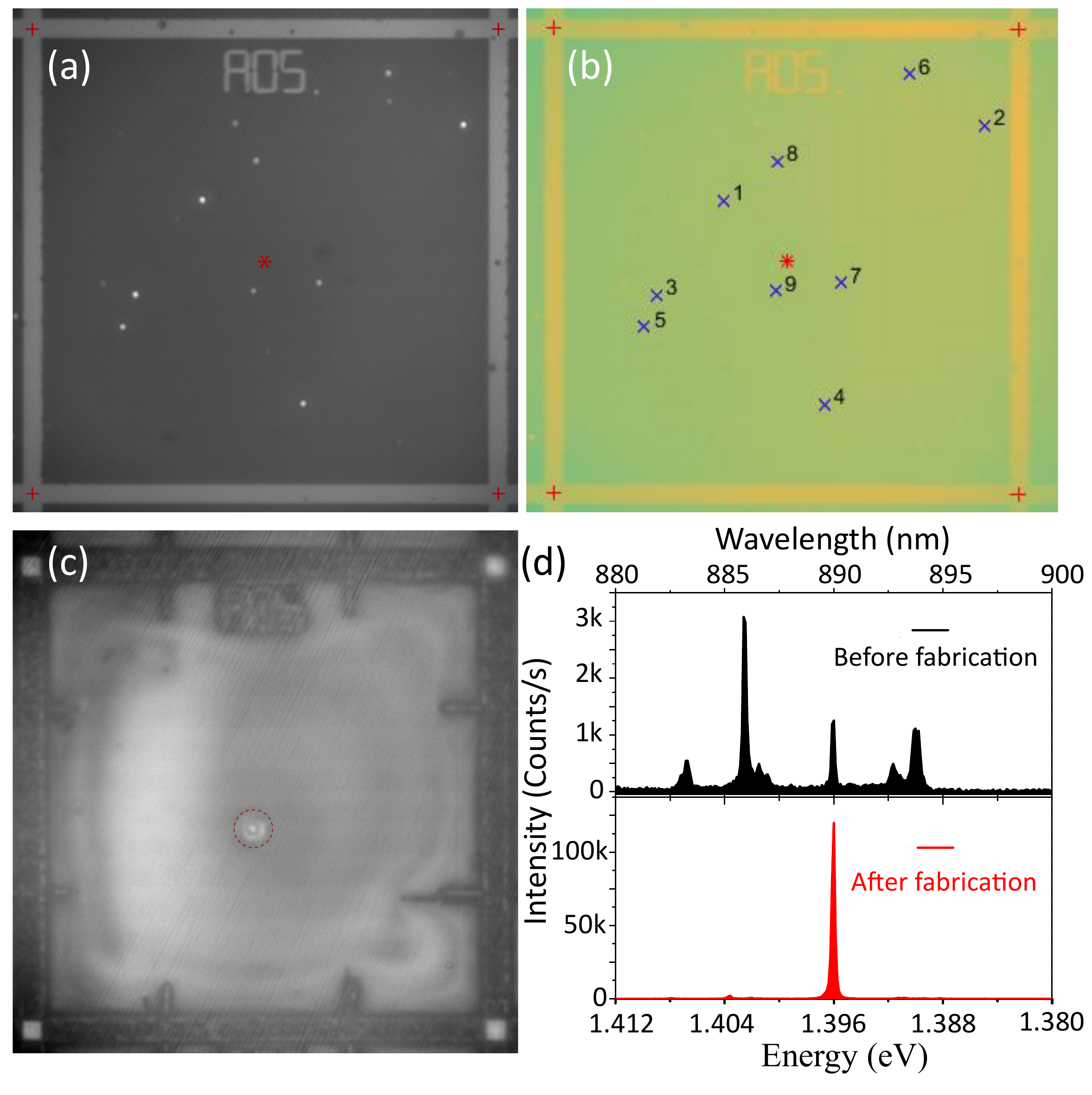}
\caption{(a) Photoluminescence image for quantum dot position extraction. The four alignment mark centers (center-to-center separation of 52~$\mu$m in both the horizontal and vertical directions) and center of the field are denoted by the red crosses and star, respectively. (b) Processed image in which individual QDs are identified and numbered. (c) Photoluminescence image of a micropillar with a single quantum dot in the center, produced by the QD positioning technique and subsequent fabrication using aligned electron-beam lithography. (d) Spectra of QD 9 before (above) and after (below) fabrication of the micropillar.}
\label{Fig2}
\end{figure}

With the photoluminescence imaging setup optimized for work with our planar DBR samples, we proceed to use it in the deterministic fabrication of micropillar single-photon sources.  Fig.~\ref{Fig2}(a) shows a photoluminescence image from the portion of the sample at 4~K that we focus on in this work, acquired in a single shot with a 1~s integration time.  By applying a combination of a maximum likelihood estimation for localizing the QD emission and a cross-correlation method for determining the alignment mark centers \cite{Liu2016}, the nine brightest QDs within a set of alignment marks are identified with sub-5 nm spatial accuracy (one standard deviation value), and numbered from 1 to 9 based on their brightness, as shown Fig.~\ref{Fig2}(b). The number of QDs to be considered for fabrication is further reduced by measuring the micro-photoluminescence ($\mu$PL) spectrum of each dot. We only make micropillar devices with the QDs whose exciton emission wavelength is within the resonance of the planar cavity, in order to match the frequency of QDs to the micropillars that are to be etched from the planar cavity. For example, among the nine QDs within the alignment marks defining field A05 in Fig.~\ref{Fig2}(a)-(b), only QD 9 has an exciton emission around the planar cavity resonance $\approx$890 nm (see the spectrum in the upper panel of Fig.~\ref{Fig2}(d)).

After spatial position extraction and spectral selection processes have been applied to the single QDs, the sample is taken out of the cryostat for further micropillar fabrication. Circular micropillar cavities are defined in spin-coated polymethyl methacrylate (PMMA) via a second e-beam lithography step. The pillars are aligned to the previously identified QDs, and ideally should only host a single emitter. The diameter of the pillars is adjusted to match the resonance condition at the lowest possible measurement temperatures. By depositing a hard mask (BaF/Cr), the pillars are transferred into the epitaxial structure via electron-cyclotron-resonance reactive-ion-etching. In the terminal step, the sample is planarized by a transparent polymer to stabilize the pillars and protect them from sidewall oxidation, and the hard mask is fully removed in an ultrasonic bath. Here, we note that a full removal of the hard mask is crucial to allow for complication-free resonant studies, since any remaining metal on top of the surface would create an unacceptable amount of stray light from the pump laser.

After fabrication of the micropillars, we have verified the successful alignment of the e-beam lithography with respect to the QD positioning in our imaging setup. With the fluorescence images of micropillar devices, it is possible to visualize that 40 out of 40 selected pillars host a single quantum emitter which is well-centered in the post. Fig.~\ref{Fig2}(c) shows a representative fluorescence image for one of the final devices, in which QD 9 sits in the center of a 2 $\mu$m diameter micropillar highlighted by the red circle. A direct comparison between the spectra taken at the same excitation power before and after fabrication shows that the emission from the targeted exciton has been greatly enhanced by the micropillar, as shown in Fig.~\ref{Fig2}(d).

\section{Single-photon source performance}

\begin{figure*}
\includegraphics[width=\linewidth]{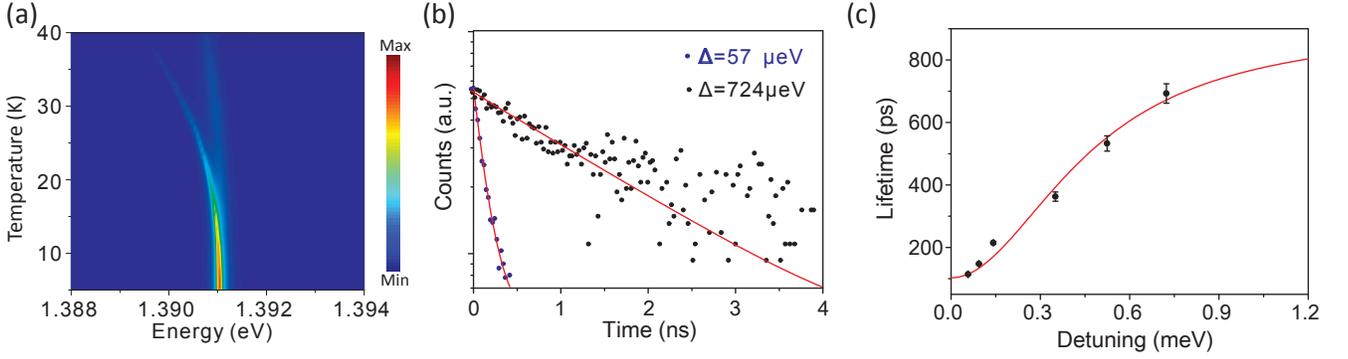}
\caption{(a) Temperature-dependent spectra of a micropillar with a diameter of $d\approx2~\mu$m under above bandgap excitation. A strong enhancement of the emission at spectral resonance due to the Purcell effect is observed. (b) Time resolved measurements on and near resonance, revealing a dramatic reduction of the radiative lifetime. (c) A fit to the QD lifetime as a function of QD-cavity detuning yields a Purcell factor of $F_P=7.8\pm1.5$. The error bars in the lifetime data are determined from fitting a mono-exponential decay to the time-resolved measurement, and are a one standard deviation value. The uncertainty in the Purcell factor represents a one standard deviation value, and is estimated from a least squares fit to the data (solid red line in (c)) according to Eq.~\ref{eqn:Purcell} in the main text.}
\label{Fig3}
\end{figure*}

Now, we demonstrate that our device indeed is capable of being operated as a competitive single photon source. We investigate a device with a diameter of $d=2~\mu$m and the extracted $Q=~4477\pm28$ (one standard deviation uncertainty from a Lorentzian fit to the data). Figure~\ref{Fig3}(a) shows a contour plot of spectra recorded from this device at various sample temperatures under non-resonant excitation. As we increase the sample temperature, the single QD is subject to a spectral redshift, and spectral resonance with the cavity mode is no longer maintained. This detuning of the QD with respect to the cavity comes along with a significant reduction of the emitted light due to reduced coupling to the guided resonant mode.
To determine the Purcell enhancement of the system, the device is operated under pulsed resonance fluorescence conditions. We employed polarization filtering in this experiment, analogous to previous works \cite{He2013}. The decay time was recorded via a fast avalanche photon diode, triggered by the resonant laser. The decay of the emission signal as a function of the time delay for the resonant as well as the far detuned (1 meV) case is shown in Fig.~\ref{Fig3}(b). From mono-exponentially decaying signals, we can extract the characteristic $T_1$ time of the QD, which is reduced to 100 ps on resonance due to the Purcell effect. To further quantify the coupling between QD and cavity, we have carried out this experiment for 6 different detunings and plotted the resulting $T_1$ times as a function of the spectral QD-cavity detuning (Fig.~\ref{Fig3}(c)). This allows us to accurately determine the Purcell factor of our QD-cavity device by fitting the lifetime as a function of the detuning with the formula:

\begin{align}
\tau(\Delta)=\frac{F_{P,Max.}}{(F_{P,Max.}*\delta+1)}*\frac{\hbar*\varepsilon_0*V_0}{2*Q*\mu_{12}^2},
\label{eqn:Purcell}
\end{align}


\noindent where $\delta=\xi^2*\frac{\Delta\omega_c^2}{4*(\Delta^2)+\Delta\omega_c^2}$, $\xi$ represents the orientation mismatch between the local cavity field and the dipole moment of the QD, $\Delta$ is the detuning, $\mu_{12}$ is the dipole moment of the radiative transition, and $\Delta\omega_c$ is the linewidth of the cavity mode.

\begin{figure}
\includegraphics[width=\linewidth]{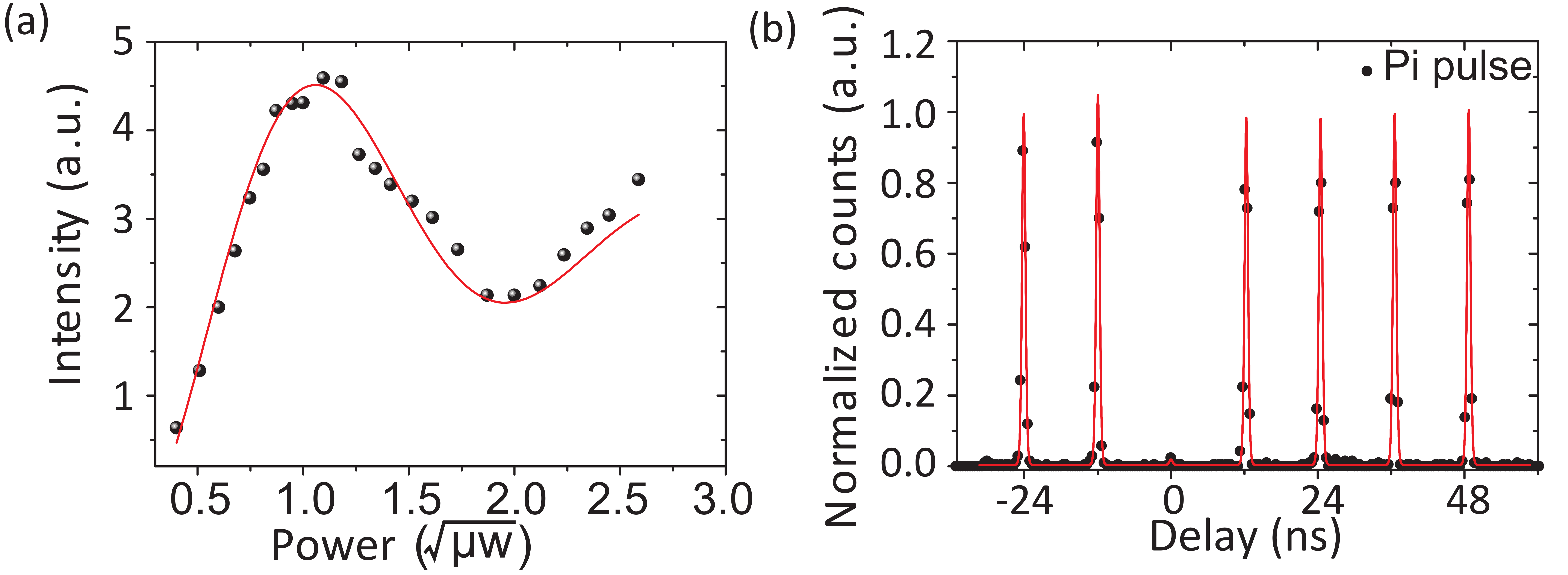}
\caption{(a) Measured count rate on the spectrometer versus the pulse area of the driving laser field for spectral resonance between the QD and cavity mode.  (b) 2$^{nd}$ order autocorrelation histogram for pulsed resonant excitation with a $\pi$-pulse. We extract a $g^{(2)}$-value as low as $g^{(2)}(0)=0.015\pm0.009$.}
\label{Fig4}
\end{figure}

For our device, we observe a maximum Purcell factor as large as $7.8 \pm 1.5$, based on the assumption that suppression of spontaneous emission off resonance is insignificant in our case. The theoretical maximum\cite{Barnes2002} of $F_{P,Max.}=\frac{3Q(\lambda/n)^3}{4\pi^2V_M}=9.73$ for a micropillar with a diameter of 2 $\mu$m and a Q-factor of 4477. Here, $n$ is the refractive index of the cavity material ($n=3.6$ for GaAs) and $V_M$ is the mode volume. For our estimation of the maximal Purcell enhancement we used the value reported in \cite{Boeckler2008} and scaled it with the volume of the micropillar. As we compare this number with the maximum Purcell factor in theory, we observe a slight reduction in our case from approximately 10 to 7.8. This can be a result of structural imperfections in the photonic device which causes increased scattering losses into non-resonant emission channels.

\begin{figure*}
\includegraphics[width=0.65\linewidth]{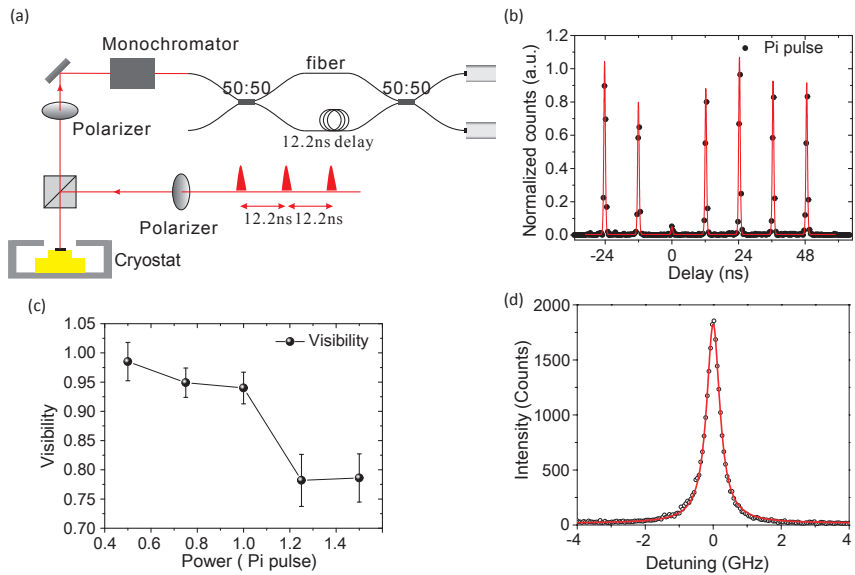}
\caption{(a) Unbalanced Mach-Zehnder interferometers with a path difference of 12.2~ns, used in characterizing the indistinguishability of single photons emitted by the QD micropillar source. (b) Histogram of two-photon interference with the 12.2~ns delay time. Photons with parallel polarization are prepared here. The histogram was fitted by a sum of 7 two-sided exponential functions, each convoluted with a Gaussian distribution. (c) Extracted photon indistinguishability as a function of pump power. An obvious power-related interference visibility decrease is revealed. (d) Measurement of the emission linewidth when the QD is on resonance with the micropillar mode and under continuous-wave, resonant laser excitation.  A linewidth of 473~MHz $\pm$ 3.0~MHz is extracted from a Lorentzian fit to the data.}
\label{Fig5}
\end{figure*}

The strictly resonant excitation of our QD allows us to establish full inversion of our effective two-level system. In a power-dependent study (Fig.~\ref{Fig4}(a)) we show the characteristic Rabi-oscillation behavior as a function of the square root of the pump power, which is the key signature for the pulsed resonant driving of the two level system. The red curve in Fig.~\ref{Fig4}(a) shows the sinusoidal damped behavior according to the exciton-phonon coupling model. For $\pi$-pulse excitation, we observe count rates on the spectrometer up to $\approx$ 130~000 counts per second. In order to extract the overall device efficiency of our QD-micropillar device from this measurement, we carefully calibrated our setup, revealing a setup efficiency for the detected linear polarization of $\eta_{Setup}=(0.325\pm0.01)~\%$, where the uncertainty is due to fluctuations in the power measurement and represents one standard deviation. Therefore, our device yields an overall extraction efficiency of $\eta=(49\pm4)~\%$. Because the QD energy is still red-shifted by 57.42~${\mu}$eV at 4.3~K, our extraction efficiency is promising when compared with the maximum extraction efficiency for the pillar in theory ($\eta=\frac{Q_{pillar}}{Q_{2D}}*\frac{F_{p,max}}{1+F_{p,max}}=(59.54\pm11.86)~\%$). The high brightness, and comparably large setup efficiency allows us to record quasi-background-free single photon correlation charts via a fiber-coupled Hanbury Brown and Twiss setup. Figure \ref{Fig4}(b) shows the recorded coincidence histogram for $\pi$-pulse excitation. The vanishing peak around $\tau\approx0$ ns is a clear signature of the non-classical light emission from the QD. We fit each pulse with a two-sided exponential decay convolved with a Gaussian distribution, where the width is the time resolution ($t_{Res}\approx520$ ps) of our setup. This allows us to extract a value of $g^{(2)}(0)$ via dividing the area of the central peak by the average area of the surrounding peaks, which amounts to $g^{(2)}(0)=0.015\pm0.009$, where the uncertainty is due to the variation in the surrounding peak area and represents a one standard deviation value.

Next, we test the coherence of consecutive emitted single photons from our device, which are excited with the repetition rate of the pump laser (82~MHz).  The emitted photons are coupled into an unbalanced fiber coupled Mach-Zehnder-interferometer (Fig.~\ref{Fig5}(a)). One arm of the interferometer is precisely adjusted to compensate for the delay (12.2~ns) between the two photons. If the early photon takes the long arm of the interferometer and the late photon the short path, both meet each other at the second 50/50 beam splitter where they can interfere if they are indistinguishable. Figure \ref{Fig5}(b) shows the measured coincidence histograms for parallel polarization of the photons for driving the system with a $\pi$-pulse. The suppression of the central peak in Fig. \ref{Fig5}(b) is a clear proof that our device emits highly coherent photons on demand. In order to extract the degree of indistinguishability, we slightly adjust the attenuation in the interferometer to balance the counts before the second 50/50 fiber beam splitter (R/T$ \approx $1) and each histogram in Fig.\ref{Fig5}(b) was fitted by a sum of 7 two-sided exponential functions convolved with a Gaussian distribution. Therefore, we could extract the raw two-photon interference visibility with the function $\upsilon_{raw}=1-\frac{A_{\parallel}}{A_{\perp}}=(91.1\pm1.9)~\%$, where the $A_{\parallel}$ was expressed by the fitted histogram area at 0 delay divided by the average area of the 4 other peaks at -24.4~ns, 24.4~ns, 36.6~ns, 48.8~ns and $A_{\perp}=0.5$. The uncertainty is a one standard deviaton value and is due to the variation in surrounding peak areas.  When taking into account the $g^{(2)}(0)$ value, we can get a corrected interference visibility of $\upsilon_{corrected}=(93.99\pm2.71)~\%$. By reducing the excitation power to correspond to a $\pi/2$ pulse, excitation-induced dephasing in our system is minimized, and near-unity indistinguishability $(98.5\pm3.2)~\%$ is fully restored, as can be seen from Fig. \ref{Fig5}(c). Finally, measurement of the QD linewidth (Fig.~\ref{Fig5}(d)) by a scanning-Fabry Perot cavity under continuous-wave, resonant driving yields a linewidth of (473 $\pm$ 3.0)~MHz, where the uncertainty is a one standard deviation value obtained from a Lorentzian fit to the data.  Considering the Purcell-enhanced radiative lifetime of the QD under the detuning conditions for this measurement, this linewidth (and corresponding coherence time $T_{2}\approx680~$ps) is consistent with near-unity degree of indistinguishability we measure.

In conclusion, we have discussed the implementation of a deterministic quantum dot -based single photon source. We apply quantum dot imaging to create our device, which is, in principle, a technology that can be used to significantly scale up the number of fabricated devices with high throughput. Our device features a Purcell factor of $F_P=7.8$, which is highly beneficial for the emission of single photons close to the Fourier limit. In a power-dependent study, we demonstrate unity indistinguishability of the emitter resonance fluorescence photons at $\pi/2$ conditions, which are only slightly compromised to a visibility of $94 ~\%$ as the two-level system is fully inverted. We believe that QD imaging represents a superior technology for the integration of single QDs into photonic architectures, as it is highly compatible with any kind of device geometry. It furthermore benefits from its almost scalable nature, induced by the one-shot identification of single QD positions.

\section{Supplementary Material}
\subsection{Sample growth}
The micro-cavity sample was grown by means of molecular beam epitaxy (MBE) on a GaAs (001) substrate. The bottom (top) distributed Bragg reflector (DBR) consists of 25.5 (15) pairs of quarter-wavelength thick layers of AlAs and GaAs. Self-assembled InAs quantum dots with a density of about $10^9$ cm$^{-2}$ are centered in a $\lambda$ thick GaAs cavity with a silicon delta doped layer 10 nm underneath the QDs. To shift the emission wavelength of the QDs to the 900 nm range, we performed an \textit{in situ} partial capping and annealing step during the MBE growth.

\vspace{0.1in}

\subsection{Supplementary Material: Optical measurements and Hong-Ou-Mandel interference}
The setup for the strictly resonant excitation of the deterministic micropillar sample consists of a single mode (SM) fiber-coupled, linear polarized Ti:Sapphire laser (repetition rate $82$ MHz, pulse length $\tau\approx1.3$ ps) which is coupled into the optical beam path via a $8/92$ pellicle beamsplitter. The laser is focused onto the sample with a microscope objective (NA=0.42) which also collects the emitted single photons. The QD signal is filtered with a second linear polarizer and a SM fiber. After spectrally filtering the signal on a monochromator, we can analyze it on a CCD or couple it into a SM fiber. The single photon statistics are measured via a fiber-coupled Hanbury-Brown and Twiss setup. For the two-photon interference experiment, the spectrally filtered single photons are then coupled into a fiber--coupled, unbalanced Mach-Zehnder-interferometer, where the two arms exactly compensate the delay of the exciting laser. We measure the second order autocorrelation of this interferometer via two single-photon avalanche diodes at the exit ports of the second $50/50$ beamsplitter.

\section*{Funding Information}
We acknowledge financial support by the State of Bavaria and the German Ministry of Education and Research (BMBF) within the projects Q.com-H. Support by the DFG within the project SCHN1376-2.1 (DACH) is greatfully acknowledged.
Y.-M. H. acknowledges support from the Sino-German (CSC-DAAD) Postdoc Scholarship Program.
J.-L acknowledges support by the Cooperative Research Agreement between the University of Maryland and NIST-CNST, Award 70NANB10H193, National Natural Science Foundation of China (grant no. 11304102) and the Ministry of Science and Technology of China (grant no. 2016YFA0301300).

\section*{Acknowledgements}

The authors would like to thank A. Wolf for assistance during the lithography.




%

\end{document}